\documentclass[10pt,conference]{IEEEtran}
\usepackage{graphicx}
\usepackage{epsfig}
\usepackage{amsmath,amsfonts,dsfont,amssymb,stmaryrd}
\newtheorem{definition}{Definition}
\newtheorem{thm}{Theorem}
\newtheorem{prop}{Proposition}
\newtheorem{lemma}{Lemma}
\newtheorem{conj}{Conjecture}

\newcommand\independent{\protect\mathpalette{\protect\independent}{\perp}} 
\def\independent#1#2{\mathrel{\rlap{$#1#2$}\mkern2mu{#1#2}}} 

\newcommand{\mR}{\mathbb{R}}

\newcommand{\N}{\mathcal{N}}
\newcommand{\e}{\varepsilon}
\newcommand{\wtil}{\widetilde}

\newcommand{\Gf}{\Gamma^{(+)}}
\newcommand{\Gb}{\Gamma^{(-)}}

\newcommand{\eqe}{\stackrel{\cdot}{=}}
\newcommand{\iid}{\stackrel{\mathrm{iid}}{\sim}}

\newcommand{\E}{\mathbb{E}}

\begin{document}

\title{
A Coordinate System for Gaussian Networks}

\author{\authorblockN{Emmanuel A. Abbe}
\authorblockA{IPG, EPFL \\
Lausanne, 1015 CH \\
Email: emmanuel.abbe@epfl.ch}
\and
\authorblockN{Lizhong Zheng}
\authorblockA{RLE, MIT\\
Cambridge, MA 02139\\
Email: lizhong@mit.edu}
}

\maketitle



\begin{abstract}
This paper studies network information theory problems where the external noise is Gaussian distributed. In particular, the Gaussian broadcast channel with coherent fading and the Gaussian interference channel are investigated. It is shown that in these problems, non-Gaussian code ensembles can achieve higher rates than the Gaussian ones. It is also shown that the strong Shamai-Laroia conjecture on the Gaussian ISI channel does not hold.
In order to analyze non-Gaussian code ensembles over Gaussian networks, a geometrical tool using the Hermite polynomials is proposed. This tool provides a coordinate system to analyze a class of non-Gaussian input distributions that are invariant over Gaussian networks. 

\end{abstract}

\section{Introduction}\label{intro}
Let a memoryless additive white Gaussian noise (AWGN) channel be described by $Y=X+Z$, where $Z \sim \N (0,v)$ is independent of $X$. If the input is imposed an average power constraint given by $\E X^2 \leq p$, the input distribution maximizing the mutual information is Gaussian. This is due to the fact that under second moment constraint, the Gaussian distribution maximizes the entropy, hence 
\begin{align}
\arg \max_{X : \E X^2 = p} h(X + Z) \sim \N (0,p). \label{r1}
\end{align}
On the other hand, if we use a Gaussian input distribution, i.e., $X \sim \N (0,p)$, the worst noise that can occur, i.e., the noise minimizing the mutual information, among noises with bounded second moment, is again Gaussian distributed. This can be shown by using the {\it entropy power inequality} (EPI), cf. \cite{stam}, which reduces in this setting to
\begin{align}
\arg \min_{Z:\, h(Z)=\frac{1}{2} \log 2 \pi e v} h(X+Z) \sim \mathcal{N}(0,v) 
\end{align}
and implies 
\begin{align}
\arg \min_{Z : \E Z^2 = v} h(X + Z) - h(Z) \sim \N (0,v). \label{r2}
\end{align}
Hence, in the single-user setting, when optimizing the mutual information as above, 
a Gaussian input is the best input for a Gaussian noise and a Gaussian noise is the worst noise for a Gaussian input. This provides a game equilibrium between user and nature, as defined in \cite{cover}, p. 263. With these results, many problems in information theory dealing with Gaussian noise can be solved. However, in Gaussian networks, that is, in multi-user information theory problems where the external noise is Gaussian distributed, several new phenomena make the search for the optimal input ensemble more complex. 
Besides for some specific cases of Gaussian networks, we still do not know how interference should be treated in general. 
Let us consider two users interfering on each other in addition to suffering from Gaussian external noise and say that the receivers treat interference as noise. Then, if the first user has drawn its code from a Gaussian ensemble, the second user faces a frustration phenomenon: using a Gaussian ensemble maximizes its mutual information but minimizes the mutual information of the first user. It is an open problem to find the optimal input distributions for this problem. This is one illustration of the complications appearing in the network setting. Another example is regarding the treatment of the fading. Over a single-user AWGN channel, whether the fading is deterministic or random, but known at the receiver, does not affect the optimal input distribution. From \eqref{r1}, it is clear that maximizing $I(X;X+Z)$ or $I(X;HX+Z|H)$ under an average power constraint is achieved by a Gaussian input. However, the situation is different if we consider a Gaussian broadcast channel (BC). When there is a  deterministic fading, using \eqref{r1} and \eqref{r2}, the optimal input distribution can be shown to be Gaussian. However, it has been an open problem to show whether Gaussian inputs are optimal or not for a Gaussian BC with a random fading known at the receiver, even if the fading is such that it is a degraded BC. 

A reason for these open questions in the network information theoretic framework, is that Gaussian ensembles are roughly the only ensembles that can be analyzed over Gaussian networks, as non-Gaussian ensembles have left most problems in an intractable form. 
In this paper, a novel technique is developed to analyze a class of non-Gaussian input distributions over Gaussian noise channels. This technique is efficient to analyze the {\it competitive situations} occurring in the network problems described below.
It allows in particular to find certain non-Gaussian ensembles that outperform Gaussian ones on a Gaussian BC with coherent fading channel, a two user interference channel, and it allows to disprove the strong Shamai-Laroia conjecture on the Gaussian intersymbol interference channel. This tool provides a new insight on Gaussian networks and confirms that non-Gaussian ensembles do have a role to play in these networks. 
We now introduce with more details the notion of competitive situations.
 
\subsection{Competitive Situations}
\subsubsection{Fading Broadcast Channel}\label{bc} 
Consider a degraded Gaussian BC with coherent memoryless fading, where the fading is indeed the same for both receivers, i.e.
\begin{eqnarray*}
Y_1 &=& H X + Z_1,  \\
Y_2 &=& H X + Z_2
\end{eqnarray*}  
but $Z_1 \sim \N(0,v_1)$ and $Z_2 \sim \N(0,v_2)$, with $v_1 < v_2$. The input $X$ is imposed a power constraint denoted by $p$. 
Because the fading is coherent, each receiver also knows the realization of $H$, at each channel use. The fading and the noises are memoryless (iid) processes.
Since this is a degraded broadcast channel, the capacity region is given by all rate pairs 
$$(I(X;Y_1|U,H) ,I(U;Y_2|H))$$ with $U - X- (Y_1,Y_2)$. 
The optimal input distributions, i.e., the distributions of $(U,X)$ achieving the capacity region boundary, are given by the following optimization, where $\mu \in \mR$,
\begin{align}
\arg\max_{(U,X): \,U - X- (Y_1,Y_2) \atop{\E X^2 \leq p} } I(X;Y_1|U,H) + \mu I(U;Y_2|H). \label{fb}
\end{align}
Note that the objective function in the above maximization is given by 
$$h(Y_1|U,H) - h(Z_1) + \mu h(Y_2|H) - \mu h(Y_2|U,H).$$
Now, each term in this expression is individually maximized by a Gaussian distribution for $U$ and $X$, but these terms are combined with different signs,
so there is a competitive situation and the maximizer is not obvious. When $\mu \leq1$, one can show that Gaussian distributions are optimal. Also, if $H$ is compactly supported, and if $v$ is small enough as to make the support of $H$ and $1/v H$ non overlapping, the optimal distribution of $(U,X)$ is jointly Gaussian (cf. \cite{tunsha}). However, in general the optimal distribution is unknown. We do not know if it because we need more theorems, or if it is really that with fading, non-Gaussian codes can actually perform better than the Gaussian ones.

\subsubsection{Interference Channel}\label{ic}
We consider the symmetric memoryless interference channel (IC) with two users and white Gaussian noise. 
The average power is denoted by $p$, the interference coefficients by $a$, and the respective noise by $Z_1$ and $Z_2$ (independent standard Gaussian). We define the following expression
\begin{align}
&m S_{a,p}(X_1^m,X_2^m)\label{optint} \\ & = I(X_1^m;X_1^m+a X_2^m+Z_1^m) + I(X_2^m;X_2^m+a X_1^m+Z_2^m)  \notag \\
&=h(X_1^m+a X_2^m + Z_1^m) - h(a X_2^m + Z_1^m) \notag \\
& \quad + h(X_2^m+a X_1^m + Z_2^m) - h(a X_1^m + Z_2^m) \notag ,
\end{align}  
where $X_1^m$ and $X_2^m$ are independent random vectors of dimension $m$ with a covariance having a trace bounded by $mp$ and $Z_i^m$, $i=1,2$, are iid standard Gaussian. For any dimension $m$ and any distributions of $X_1^m$ and $X_2^m$,  $S_{a,p}(X_1^m,X_2^m)$ is a lower bound to the sum-capacity. Moreover, it is tight by taking $m$ arbitrarily large and $X_1^m$ and $X_2^m$ maximizing \eqref{optint}. 
Now, a similar competitive situation as for the fading broadcast problem takes place: 
Gaussian distributions maximize each entropy term, but these terms are combined with different signs.
Would we then prefer to take $X_1$ and $X_2$ Gaussian or not? This should depend on the value of $a$. 
If $a=0$, we have two parallel AWGN channels with no interference, and Gaussian inputs are optimal. We can then expect that this might still hold for small values of $a$. 
It has been proved recently in \cite{veer,khan,kram}, that the sum-capacity is achieved by treating interference as noise and with iid Gaussian inputs, as long as $p a^3 + a- 1/2 \leq 0$. Hence, in this regime, the iid Gaussian distribution maximizes \eqref{optint} for any $m$.
But if $a$ is above that threshold and below $1$, the problem is open. 

Let us now review the notion of ``treating interference as noise''.
For each user, we say that the decoder is {\it treating interference as noise}, if it does not require the knowledge of the other user's code book. However, we allow such decoders to have the knowledge of the distribution, under which the other user's code book may be drawn. This is for example necessary to construct a sum-capacity achieving code\footnote{in a low interference regime} in \cite{veer,khan,kram}, where the decoder of each user treats interference as noise but uses the fact that the other user's code book is drawn from an iid Gaussian distribution. But, if we allow this distribution to be of arbitrarily large dimension $m$ in our definition of treating interference as noise, we can get a misleading definition. Indeed, no matter what $a$ is, if we take $m$ large enough and a distribution of $X_1^m$, $X_2^m$ maximizing \eqref{optint}, we can achieve rates arbitrarily close to the sum-capacity, yet, formally treating interference as noise. The problem is that the maximizing distributions in \eqref{optint} may not be iid for an arbitrary $a$, and knowing it at the receiver can be as much information as knowing the other user's code book (for example, if the distribution is the uniform distribution over a code book of small error probability). Hence, one has to be careful when taking $m$ large. In this paper, we will only work with situations that are not ambiguous with respect to our definition of treating interference as noise. 
It is indeed an interesting problem to discuss what kind of $m$-dimensional distributions would capture the meaning of treating interference as noise that we want. 
This also points out that studying the maximizers of \eqref{optint} relates to studying the concept of treating interference as noise or information. 
Since for any chosen distributions of the inputs we can achieve \eqref{optint}, the maximizers of \eqref{optint} must have a different structure when $a$ grows. For $a$ small enough, iid Gaussian are maximizing distributions, but for $a\geq 1$, since we do not want to treat interference as noise, the maximizing distributions must have a ``heavy structure'', whose characterization requires as much information as giving the entire code book. 
This underlines that an encoder can be drawn from a distribution which does not maximize \eqref{optint} for any value of $m$, but yet, a decoder may exist in order to have a capacity achieving code. This happens if $a \geq 1$, iid Gaussian inputs will achieve the sum-capacity if the receiver decodes the message of both users (one can show that the problem is equivalent to having two MAC's). However, if $a\geq 1$, the iid Gaussian distribution does not maximize $S_{a,p}(X_1,X_2)$ (for the dimension 1, hence for arbitrary dimensions). 

In any cases, if the Gaussian distribution does not maximize \eqref{optint} for the dimension 1, it means that iid Gaussian inputs and treating interference as noise is not capacity achieving, since a code which treats interference as noise and whose encoder is drawn from a distribution can be capacity achieving only if the encoder is drawn from a distribution maximizing \eqref{optint}.
Hence, understanding better how to resolve the competitive situation of optimizing \eqref{optint} is a consequent problem for the interference channel.

\subsection{ISI channel and strong Shamai-Laroia Conjecture}

\begin{conj}\label{sl}
Let $h,p,v \in \mR_+$, $X_1^G \sim \N(0,p)$ and $Z \sim \N(0,v)$ (independent of $X_1^G$).  For all $X, X_1$ i.i.d. with mean 0 and variance $p$, we have
\begin{align}
I(X; X + h X_1^G + Z)  \leq I(X; X + h X_1 + Z) .
\end{align}
\end{conj}

This conjecture has been brought to our attention by Shlomo Shamai (Shitz), who referred to the strong conjecture for a slightly more general statement, where an arbitrary memory for the interference term is allowed, i.e., where $\sum_{i=1}^{n} h_i X_i$ stands for $h X_1$. The strong conjecture then claims that picking all $X_i$'s Gaussian gives a minimizer. 
However, we will show that even for the memory one case, the conjecture does not hold.  
The weak conjecture, also referred to as the Shamai-Laroia conjecture, corresponds to a specific choice of the $h_i$'s, which arises when using an MMSE decision feedback equalizer on a Gaussian noise ISI channel, cf. \cite{sl}. This conjecture is investigated in a work in progress. \\

There are many other examples in network information theory where such competitive situations occur. Our goal in this paper is to explore the degree of freedom provided by non-Gaussian input distributions. We show that the neighborhood of Gaussian distributions can be parametrized in a specific way, 
as to simplify greatly the computations arising in competitive situations.
We will be able to precisely quantify how much a certain amount of non-Gaussianness, which we will characterize by means of the Hermite polynomials, affects or helps us in maximizing the competitive entropic functional of previously mentioned problems. 


\section{Problem Statement}
\subsection{Fading BC}
For the fading BC problem problem described in \ref{bc}, we want to determine if/when the distribution of $(U,X)$ maximizing \eqref{fb} is Gaussian or not.
\subsection{IC} 
For the interference channel problem described in \ref{ic}, we know from 
\cite{veer,khan,kram} that treating interference as noise and using iid Gaussian inputs is optimal when $p a^3 + a- 1/2 > 0$. 
We question when this coding scheme is no longer optimal. More generally, we want to analyze the maximizers of \eqref{optint}.

We distinguish the implication of such a threshold in both the synchronized and asynchronized users setting, as there will be an interesting distinction between these two cases. We recall how the synch and asynch settings are defined here. In the synch setting, each user of the IC sends their code words of a common block length $n$ simultaneously, i.e., at time 1, they both send the first component of their code word, at time 2 the second component, etc. In the asynch setting, each user is still using code words of the same block length $n$, however, there might be a shift between the time at which the first and second users start sending their code words. We denote this shift by $\tau$, and assume w.l.o.g. that $0 \leq \tau \leq n$. In the totally asynch setting, we assume that $\tau$ is drawn uniformly at random within $\{0,\ldots, n\}$. We may also distinguish the cases where $\tau$ is not known at the transmitter but at the receiver, and when $\tau$ is not known at both. 
Note that if iid input distributions are used to draw the code books, and interference is treated as noise, whether the users are synch or asynch is not affecting the rate achievability\footnote{hence, 
\eqref{optint} with an iid distribution for $X_1$ and $X_2$ can still be defined for the totally asynch IC}.  However, if the users want to time-share over the channel uses, such as to fully avoid their interference, they will need synchronization. 
\begin{definition}\label{ts}
Time sharing over a block length $n$ (assumed to be even) with Gaussian inputs refers to using $X_1$ Gaussian with covariance $2P\hat{I}_{n/2}$ and $X_2$ Gaussian with covariance $2P  \hat{I}_{n/2}^c$, where $\hat{I}_{n/2}$ is a diagonal matrix with $n/2$ 1's and 0's, and $\hat{I}_{n/2}^c$ flips the 1's and 0's on the diagonal. 
\end{definition}

\subsection{ISI Channel and Strong Shamai-Laroia Conjecture}
We want to determine whether conjecture \ref{sl} holds or not.

\subsection{General Problem}
Our more general goal is to understand better the problem posed by any competitive situations. For this purpose, we formulate the following mathematical problem. 

We start by changing the notation and rewrite \eqref{r1} and \eqref{r2} as
\begin{align}
& \arg \max_{f :\,  m_2(f) = p} h(f \star g_v) = g_p \label{p1} \\
& \arg \min_{f:\, m_2(f) = p} h(f \star g_v) - h(f)=g_p \label{p2}
\end{align}
where $g_p$ denotes the Gaussian density with zero mean and variance $p$, and the functions $f$ are density functions on $\mR$, i.e., positive functions integrating to 1, and having a well-defined entropy and second moment $m_2(f)=\int_\mR x^2 f(x) dx$. \\

We consider the local geometry by looking at densities of the form 
\begin{align} 
f_\e(x)=g_p(x) (1 + \e L(x)), \quad x \in \mR,
\end{align} 
where $L:\, \mR \rightarrow \mR$ satisfies 
\begin{align} 
& \inf_{x\in \mR} L(x) > -\infty  \label{v1} \\
& \int_\mR L(x)g_p(x) dx=0. \label{v2}
\end{align} 
With these two constraints on $L$, $f_\e$ is a valid density for $\e$ sufficiently small. It is a perturbed Gaussian density, in a ``direction'' $L$.
Observe that,
\begin{align} 
& m_1(f_\e)=0 \quad \text{iff} \quad M_1(L)=\int_\mR x L(x)g_p(x) dx =0 \label{c1} \\
& m_2(f_\e)=p\quad \text{iff} \quad M_2(L)=\int_\mR x^2 L(x)g_p(x) dx=0 \label{c2}.
\end{align} 

We are now interested in analyzing how these perturbations affect the output distributions through an AWGN channel. Note that, if the input distribution is a Gaussian $g_p$ perturbed in the direction $L$, the output is a Gaussian $g_{p+v}$ perturbed in the direction $\frac{(g_p L) \star g_v }{g_{p+v}},$ since 
$$f_\e \star g_v = g_{p+v}(1 + \e \frac{(g_p L) \star g_v }{g_{p+v}}). $$
{\it Convention:} $g_p L \star g_v$ refers to $(g_p L) \star g_v$, i.e., the multiplicative operator precedes the convolution one. \\
For simplicity, let us assume in the following that the function $L$ is a polynomial satisfying \eqref{v1}, \eqref{v2}. 
\begin{lemma}\label{appro}
We have
\begin{eqnarray*}
D( f_\e ||g_p) &=& \frac{1}{2} \e^2 \| L \|_{g_p}^2 + o(\e^2) \\
D(f_\e \star g_v ||g_p \star g_v) &=& \frac{1}{2} \e^2 \| \frac{g_p L \star g_v}{g_{p+v}} \|_{g_{p+v}}^2 + o(\e^2).
\end{eqnarray*}
where $$\| L \|_{g_p}^2 = \int_\mR L^2(x) g_p(x) dx.$$
\end{lemma}
\vspace{.5cm}
Moreover, note that for any density $f$, if $m_1(f)=a$ and $m_2(f)=p+a^2$, we have 
\begin{eqnarray}
h(f) & =& h(g_{a,p}) - D(f || g_{a,p}). \label{mean-var}
\end{eqnarray}
Hence, the extremal entropic results of \eqref{p1} and \eqref{p2} are locally expressed as
\begin{align}
& \arg \min_{L :\,  M_2(L) = 0} \| \frac{g_p L \star g_v}{g_{p+v}} \|_{g_{p+v}}^2 = 0 \label{q1} \\
& \arg \max_{L:\, M_2(L) = 0} \| \frac{g_p L \star g_v}{g_{p+v}} \|_{g_{p+v}}^2 - \| L \|_{g_p}^2 =0,\label{q2}
\end{align}
where 0 denotes here the zero function. If \eqref{q1} is obvious, \eqref{q2} requires a proof which will be done in section \ref{proofs}. 
Let us define the following mapping, 
\begin{align}
\Gf: \,\,\,L \in L_2(g_p) \mapsto \frac{g_p L \star g_v }{g_{p+v}} \in L_2(g_{p+v};\mR),
\end{align}
where $L_2(g_p)$ denotes the space of real functions having a finite $\| \cdot \|_{g_p}$ norm. This linear mapping gives, for a given perturbed direction $L$ of a Gaussian input $g_p$, the resulting perturbed direction of the output through additive Gaussian noise $g_v$. The norm of each direction in their respective spaces, i.e., in $L_2(g_p)$ and $L_2(g_{p+v})$, gives how far from the Gaussian distribution these perturbations are (up to a scaling factor). Note that if $L$ satisfies \eqref{v1}-\eqref{v2}, so does $\Gf L $ for the measure $g_{p+v}$. The result in \eqref{q2} (worst noise case) tells us that this mapping is a contraction, but for our goal, what would be helpful is a spectral analysis of this operator, to allow more quantitative results than the extreme-case results of \eqref{q1} and \eqref{q2}. 

%
In order to do so, one can express $\Gf$ as an operator defined and valued in the same space, namely $L_2$ with the Lebesgue measure $\lambda$, which is done by inserting the Gaussian measure in the operator argument. We then proceed to a singular function/value analysis. Formally, let $K=L \sqrt{g_p},$ which gives $ \|K\|_\lambda = \|L\|_{g_p}$, and let
\begin{align}
\Lambda: \,\,\,K \in L_2(\lambda) \mapsto \frac{\sqrt{g_p} K\star g_v }{\sqrt{g_{p+v}}} \in L_2(\lambda)
\end{align}
which gives $ \| \Gf L  \|_{g_{p+v}} = \|\Lambda K\|_\lambda.$
Denoting by $\Lambda^*$ the adjoint operator of $\Lambda$, we want to find the singular functions of $\Lambda$, i.e., the eigenfunctions $K$ of $\Lambda^* \Lambda$:
$$ \Lambda^* \Lambda K = \gamma K .$$

\section{Results}
\subsection{General Result: Local Geometry and Hermite Coordinates}
The following theorem gives the singular functions and values of the operator $\Lambda$ defined in previous section.
\begin{thm}\label{hermite1}
$$\Lambda^* \Lambda K = \gamma K , \quad K \neq 0$$
holds for each pair $$(K,\gamma) \in \{( \sqrt{g_p} H_k^{[p]}, \left(\frac{p}{p+v}\right)^k)\}_{k \geq 0},$$
where $$H_k^{[p]}(x)=\frac{1}{\sqrt{k!}} H_k(x/\sqrt{p})$$
$$H_k(x) =  (-1)^k e^{x^2/2} \frac{d^k}{d x^k} e^{-x^2/2}, \quad k \geq 0,x\in \mR.$$
\end{thm}
\vspace{.5cm}
The polynomials $H_k^{[p]}$ are the normalized Hermite polynomials (for a Gaussian distribution having variance $p$) and $\sqrt{g_p} H_k^{[p]}$ are called the Hermite functions. For any $p>0$, $\{H_k^{[p]}\}_{k\geq 0}$ is an orthonormal basis of $L_2(g_p)$, this can be found in \cite{stroock}. 
One can check that $H_1$, respectively $H_2$ perturb a Gaussian distribution into another Gaussian distribution, with a different first moment, respectively second moment. For $k\geq 3$, the $H_k$ perturbations are not modifying the first two moments and are moving away from Gaussian distributions.
Since $H_0^{[p]}=1$, the orthogonality property implies that $H_k^{[p]}$ satisfies \eqref{v2} for any $k > 0$. However, it is formally only for even values of $k$ that \eqref{c2} is verified (although we will see in section \ref{proofs} that essentially any $k$ can be considered in our problems). The following result contains the property of Hermite polynomials mostly used in our problems, and expresses Theorem \ref{hermite1} with the Gaussian measures. 

The following result contains the property of Hermite polynomials mostly used in our problems, and expresses Proposition \ref{hermite1} with the Gaussian measures. 

\begin{thm}\label{hermite2}
\begin{align}
&\Gf H_k^{[p]}= \frac{g_p H_k^{[p]} \star g_v}{g_{p+v}}=\left(\frac{p}{p+v}\right)^{k/2} H_k^{[p+v]}, \label{it} \\
&\Gb H_k^{[p+v]}= H_k^{[p+v]} \star g_v =\left(\frac{p}{p+v}\right)^{k/2} H_k^{[p]}. 
\end{align}
\end{thm}
\vspace{.5cm}
Last Theorem implies Theorem \ref{hermite1}, since
$$\Gb \Gf L = \gamma L \quad \Longleftrightarrow \quad  \Lambda^* \Lambda K = \gamma K $$
for $$K= L  \sqrt{g_p}.$$

{\it Comment:} the results that we have just derived are related to properties of the Ornstein-Uhlenheck process. 



{\it Summary:} 
In words, we just saw that $H_k $ is an eigenfunction of the input/output perturbation operator $\Gf$, in the sense that $\Gf H_k^{[p]} = \left(\frac{p}{p+v} \right)^{k/2} H_k^{[p+v]}$. Hence, over an additive Gaussian noise channel $g_v$, if we perturb the input $g_p$ in the direction $H_k^{[p]}$ by an amount $\e$, we will perturb the output $g_{p+v}$ in the direction $H_k^{[p+v]}$ by an amount $ \left(\frac{p}{p+v} \right)^{k/2}  \e$. Such a perturbation in $H_k$ implies that the output entropy is reduced (compared to not perturbing) by $ \left(\frac{p}{p+v} \right)^k \frac{\e^2}{2}$ (if $k \geq 3$).




\subsection{Fading BC Result}
The following result states that the capacity region of a degraded fading BC with Gaussian noise is not achieved by a Gaussian superposition code in general. 
\begin{thm}\label{ce}
Let
\begin{align*}
& Y_1 = H X + Z_1, \\ 
& Y_2 = H X + Z_2
\end{align*} 
with $X$ such that $\E X^2 \leq p$, $Z_1 \sim \N(0,v)$, $0<v<1$, $Z_2 \sim \N(0,1)$ and $H,X,Z_1,Z_2$ mutually independent. 
There exists a fading distribution and a value of $v$ for which the capacity achieving input distribution is non-Gaussian. More precisely, let $U$ be any auxiliary random variable, with $U-X-(Y_1,Y_2)$. 
Then, there exists $p,v$, a distribution of $H$ and $\mu$ such that  
\begin{align}
(U,X) \mapsto I(X;Y_1|U,H) + \mu I(U;Y_2|H) \label{map} 
\end{align}
is maximized by a non jointly Gaussian distribution. 
\end{thm}

\vspace{.2cm}
In the proof, we present a counter-example to Gaussian being optimal for $H$ binary. In order to defeat Gaussian distributions, we construct input distributions using the Hermite coordinates. The proof also gives a condition on the fading distribution and the noise variance $v$ for which a non-Gaussian distribution strictly improves on the Gaussian one.


\subsection{IC Result}

\begin{definition}
Let $$ F_k(a,p) =\lim_{\delta \searrow 0} \lim_{\e \searrow 0} \frac{2}{\e^2} \left[ S_{a,p}(X_1,X_2) - S_{a,p}(X_1^G,X_2^G) \right],$$
where $X_1^G , X_2^G \iid g_p$, $X_1 \sim g_p(1+ \e \wtil{H}_k)$ and $X_2 \sim g_p(1- \e \wtil{H}_k)$, with $\wtil{H}_k$ defined in \eqref{tild} below (as explain in section \ref{form}, $\wtil{H}_k$ is a formal modification of $H_k$ to ensure the positivity of the perturbed densities). \\
In other words, $F_k(a,p)$ represents the gain (positive or negative) of using $X_1$ perturbed along $H_k$ and $X_2$ perturbed along $-H_k$ with respect to using Gaussian distributions. 
Note that the distributions we chose for $X_1$ and $X_2$ are not the most general ones, as we could have chosen arbitrary directions spanned by the Hermite basis to perturb the Gaussian densities. However, as explained in the proof of the theorem \ref{calc}, this choice is sufficient for our purpose.

\end{definition}

\begin{thm}\label{calc}
We have for $k \geq 2$
$$F_k(a,p) = \left[ \frac{a^{2}}{p a^2+1} \right]^k - \frac{(a^k-1)^2}{(pa^2+p+1)^k}.$$
\end{thm}
\vspace{.2cm}
For any fixed $p$, the function $F_k(\cdot,p)$ has a unique positive root, below which it is negative and above which it is positive. 


\begin{thm}\label{inter}
Treating interference as noise with iid Gaussian inputs does not achieve the sum-capacity of the symmetric IC (synch or asynch) and is outperformed by $X_1 \sim g_p(1+ \e \wtil{H}_3)$ and $X_2 \sim g_p(1- \e \wtil{H}_3)$, if $F_3(a,p)>0.$
\end{thm}
\vspace{.1cm} 
This Theorem is a direct consequence of Theorem \ref{calc}. 
\begin{prop}\label{f2}
For the symmetric synch IC, time sharing improves on treating interference as noise with iid Gaussian distribution if $F_2(a,p)>0.$ 
\end{prop}
\vspace{.1cm}
We now introduce the following definition. 
\begin{definition}
Blind time sharing over a block length $n$ (assumed to be even) between two users, refers to sending non-zero power symbols only at the instances marked with a 1 in $(1,0,1,0,1,0,\ldots1,0)$ for the first user, and zero power symbols only at the instances marked with a 1 in $(1,1,\ldots,1,0,0,\ldots,0)$ for the second user. 
\end{definition}

\begin{prop}\label{b2}
For the symmetric totally asynch IC, if the receivers (but not transmitters) know the asynchronization delay, blind time sharing improves on treating interference as noise with iid Gaussian distributions if $B_2(a,p)>0,$
where $B_2(a,p) =\frac{1}{4}(\log(1+2p)+\log(1+\frac{2p}{1+2a^2 p}))-\log(1+\frac{p}{1+a^2 p}).$
If the receivers do not know the asynchronization delay, blind time sharing cannot improve on treating interference as noise with iid Gaussian distributions if $B_2(a,p)\leq 0$. 
\end{prop}

{\it How to read these results:} We have four thresholds to keep track of:
\begin{itemize}
\item $T_1(p)$ is when $p a^3 + a - \frac{1}{2} = 0$. If $a\leq T_1(p)$, we know from \cite{veer,khan,kram} that iid Gaussian inputs and treating interference as noise is sum-capacity achieving.
\item $T_2(p)$ is when $F_2(a,p)=0$. If $a > T_2(p)$, we know from Prop. \ref{f2} that, if synchronization is permitted, time sharing improves on treating interference as noise with iid Gaussian inputs. This regime matches with the so-called moderate regime defined in \cite{costa}. 
\item $T_3(p)$ is when $F_3(a,p)=0$. If $a > T_3(p)$, we know from Prop. \ref{inter} that treating interference as noise with iid non-Gaussian distributions (opposites in $H_3$) improves on the iid Gaussian ones.
\item $T_4(p)$ is when $B_2(a,p)=0$. If $a > T_4(p)$, we know from Prop. \ref{b2} that, even if the users are totally asynchronized, but if the receivers know the asynchronization delay, blind time sharing improves on treating interference as noise with iid Gaussian inputs. If the receivers do not know the delay, the threshold can only appear for larger values of $a$. 
\end{itemize}
The question is now, how are these thresholds ranked. It turns out that 
$0<T_1(p)< T_2(p)<T_3(p)<T_4(p).$ And if $p=1$, the above inequality reads as $0.424<0.605<0.680<1.031$.
This implies the following for a decoder that treats interference as noise. Since $T_2(p)<T_3(p)$, it is first better to time share than using non-Gaussian distributions along $H_3$. But this is useful only if time-sharing is permitted, i.e., for the synch IC. However, for the asynchronized IC, since $T_3(p)<T_4(p)$, we are better off using the non-Gaussian distributions along $H_3$ before a Gaussian input scheme, even with blind time-sharing, and even if the receiver could know the delay. We notice that there is still a gap between $T_1(p)$ and $T_2(p)$, and we cannot say if, in this range, iid Gaussian inputs are still optimal, or if another class of non-Gaussian inputs (far away from Gaussians) can outperform them. In \cite{calvo}, another technique (which is related to ours but not equivalent) is used to find regimes where non-Gaussian inputs can improve on Gaussian ones on the same problem that we consider here. The threshold found in \cite{calvo} is equal to 0.925 for $p=1$, which is looser than the value of 0.680 found here. 

Finally, the following interesting and curious fact has also been noticed. In theorem \ref{calc}, we require $k \geq 2$. Nevertheless, if we plug $k=1$ in the right hand side of theorem \ref{calc} and ask for this expressions to be positive, we precisely get $p a^3 + a - \frac{1}{2} > 0$, i.e., the complement range delimited by $T_1(p)$. However, the right hand side of theorem \ref{calc} for $k=1$ is not equal to $F_1(a,p)$ (this is explained in more details in the proof of theorem \ref{calc}). Indeed, it would not make sense that moving along $H_1$, which changes the mean with a fixed second moment within Gaussians, would allow us to improve on the iid Gaussian scheme. Yet, getting to the exact same condition, when working on the problem of improving on the iid Gaussian scheme, seems to be a strange coincidence.

\subsection{Strong Shamai-Laroia Conjecture}

We show in Section \ref{proofs} that conjecture \ref{sl} does not hold.
We provide counter-examples to the conjecture, pointing out that the range of $h$ for which the conjecture does not hold increases with SNR.

\section{Hermite Coding: Formalities}\label{form}
The Hermite polynomial corresponding to $k=0$ is $H_0^{[p]}=1$ and is clearly not a valid direction as it violates \eqref{v2}. Using the orthogonality property of the Hermite basis and since $H_0^{[p]}=1$, we conclude that $H_k^{[p]}$ satisfies \eqref{v2} for any $k > 0$. However, it is only for $k$ even that $H_k^{[p]}$ satisfies \eqref{v1}. On the other hand, for any $\delta >0$, we have that $H_k^{[p]} + \delta H_{4k}^{[p]}$ satisfies \eqref{v1}, whether $k$ is even or not (we chose $4k$ instead of $2k$ for reasons that will become clear later). Now, if we consider the direction $- H_k^{[p]}$, \eqref{v1} is not satisfied for both $k$ even and odd. But again, for any $\delta >0$, we have that $-H_k^{[p]} + \delta H_{4k}^{[p]}$ satisfies \eqref{v1}. 
Hence, in order to ensure \eqref{v1}, we will often work in the proofs with $ \pm H_k^{[p]} + \delta H_{4k}^{[p]}$, although it will essentially allow us to reach the performance achieved by any $\pm H_k^{[p]}$ (odd or even), since we will then take $\delta$ arbitrarily small and use continuity arguments. 

{\it Convention:} 
We drop the variance upper script in the Hermite terms whenever a Gaussian density with specified variance is perturbed, i.e., the density $g_p(1+\e H_k )$ always denotes $g_p(1+\e H_k^{[p]} )$, and $g_p H_k$ always denotes $g_p  H_k^{[p]} $, no matter what $p$ is. Same treatment is done for $\| \cdot \|_{g_p}$ and $\| \cdot \| $.

Now, in order to evaluate the entropy of a perturbation, i.e., $h(g_p(1+\e L))$, we can express it as the entropy of $h(g_p)$ minus the divergence gap, as in \eqref{mean-var}, and then use Lemma \ref{appro} for the approximation. But this is correct {\it if} $g_p(1+\e L)$ has the same first two moments as $g_p$. Hence, if $L$ contains only $H_k$'s with $k \geq 3$, the previous argument can be used. But if $L$ contains $H_1$ and/or $H_2$ terms, the situation can be different. Next Lemma describes this.   
\begin{lemma}\label{corr}
Let $\delta ,p >0$ and
\begin{eqnarray}
b \wtil{H}_k^{[p]} =
   \begin{cases}
      b(H_k^{[p]} + \delta H_{4k}^{[p]}) ,& \text{if } b \geq 0,  \\
     b(H_k^{[p]} - \delta H_{4k}^{[p]}), & \text{if } b< 0. \label{tild}
   \end{cases}
\end{eqnarray}
We have for any $\alpha_k \in \mR$, $k \geq 1$, $\e >0$
\begin{align*}
& h(g_p(1+ \e \sum_{k\geq 1} \alpha_k \wtil{H}_k))= \\
&h(g_p) -D( g_p(1+ \e \sum_{k\geq 1} \alpha_k \wtil{H}_k ) ||g_p) +\frac{\e \alpha_2}{\sqrt{2}}.
\end{align*}
\end{lemma}
Finally, when we convolve two perturbed Gaussian distributions, we get $g_a(1+\e H_j) \star g_b(1+\e H_k)= g_{a+b}+ \e [g_a H_j \star g_b +  g_a \star g_b H_k] + \e^2 g_a H_j \star g_b H_k.$ We already know from Theorem \ref{hermite2} how to describe the terms in $\e$, what we still need is to describe the term in $\e^2$. We have the following. 
\begin{lemma}\label{new}
We have $$g_a H_k^{[a]} \star g_b H_l^{[b]} = C g_{a+b} H_{k+l}^{[a+b]},$$ 
where $C$ is a constant depending only on $a,b,k$ and $l$. In particular if $k=l=1$, we have $C= \frac{\sqrt{2 a b} }{a+b} $.
\end{lemma}
\vspace{.2cm}

\section{Proofs}\label{proofs}
We start by reviewing the proof of \eqref{q2}, as it brings interesting facts. We then prove the main result.

{\it Proof of \eqref{q2}:}\\
We first assume that $f_\e$ has zero mean and variance $p$. Using the Hermite basis, we express $L$ as $L=\sum_{k \geq 3} \alpha_k H_k^{[p]}$ ($L$ must have such an expansion, since it must have a finite $L_2(g_p)$ norm, to make sense of the original expressions). Using \eqref{it}, we can then express \eqref{q2} as
\begin{eqnarray}
 \sum_{k \geq 3} \alpha_k^2 \left(\frac{p}{p+v} \right)^k -  \sum_{k \geq 3} \alpha_k^2 \label{bd}
\end{eqnarray}
which is clearly negative. Hence, we have proved that
\begin{align}
 \| \frac{g_p L \star g_v}{g_{p+v}} \|_{g_{p+v}}^2 \leq \| L \|_{g_p}^2 \label{contr} 
 \end{align} 
and \eqref{q2} is maximized by taking $L=0$. Note that we can get tighter bounds than
the one in previous inequality, indeed the tightest, holding for $H_3$, is given by
\begin{align}
\| \frac{g_p L \star g_v}{g_{p+v}} \|_{g_{p+v}}^2 \leq \left(\frac{p}{p+v} \right)^3 \| L \|_{g_p}^2 \label{tight}
\end{align}
(this clearly holds if written as a series like in \eqref{bd}).
Hence, locally the contraction property can be tightened, and locally, we have stronger EPI's, or worst noise case. Namely, if $\nu \geq \left( \frac{p}{p+v} \right)^3$, we have
\begin{align}
& \arg \min_{f:\, m_1(f)=0,m_2(f) = p} h(f \star g_v) - \nu h(f)=g_p 
\end{align}
and if $\nu < \left( \frac{p}{p+v} \right)^3$, $g_p$ is outperformed by non-Gaussian distributions. Now, if we consider the constraint $m_2(f) \leq p$, which, in particular, allows to consider $m_1(f)>0$ and $m_2(f)=p$, we get that if $\nu \geq \frac{p}{p+v}$, 
\begin{align}
& \arg \min_{f:\, m_2(f) \leq p} h(f \star g_v) - \nu h(f)=g_p 
\end{align}
and if $\nu < \frac{p}{p+v}$, $g_p$ is outperformed by $g_{p-\delta}$ for some $\delta>0$.
It would then be interesting to study if these tighter results hold in a greater generality than for the local setting.

{\it Proof of Theorem \ref{hermite2}:}\\
We want to show
\begin{align*}
& H_k^{[p+v]} \star g_v =(\frac{p}{p+v})^{k/2} H_k^{[p]}, \\
& g_p H_k^{[p]} \star g_v=(\frac{p}{p+v})^{k/2} g_{p+v} H_k^{[p+v]},
\end{align*}
which is proved by an induction on $k$, using the following properties (Appell sequence and recurrence relation) of Hermite polynomials:
\begin{eqnarray*}
\frac{\partial}{\partial x} H_{k+1}^{[p]} (x)&=& \sqrt{\frac{k+1}{p}} H_{k}^{[p]} (x) \\
\frac{\partial}{\partial x} \left( g_p (x)H_k^{[p]}(x) \right) &=&-\sqrt{\frac{k+1}{p}} g_p(x) H_{k+1}^{[p]}(x).
\end{eqnarray*}

{\it Proof of Theorem \ref{ce}:}\\
We refer to \eqref{map} as the mu-rate. Let us first consider Gaussian codes, i.e., when $(U,X)$ is jointly Gaussian, and see what mu-rate they can achieve. Without loss of generality, we can assume that $X=U+V$, with $U$ and $V$ independent and Gaussian, with respective variance $Q$ and $R$ satisfying $P=Q+R$. Then, \eqref{map} becomes 
\begin{align}
\frac{1}{2} \E \log (1+ \frac{R H^2}{v}) + \mu \frac{1}{2} \E \log \frac{1+P H^2}{1+R H^2} . \label{gauss}
\end{align}
Now, we pick a $\mu$ and look for the optimal power $R$ that must be allocated to $V$ in order to maximize the above expression. We are interested in cases for which the optimal $R$ is not at the boundary but at an extremum of \eqref{gauss}, and if the maxima is unique, the optimal $R$ is found by the first derivative check, which gives $\E \frac{ H^2}{v+RH^2} = \mu \E \frac{H^2}{1+RH^2}.$
Since we will look for $\mu$, $v$, with $R>0$, previous condition can be written as
\begin{align}
 \E \frac{R H^2}{v+RH^2} = \mu \E \frac{RH^2}{1+RH^2}. \label{optmu}
\end{align}
We now check if we can improve on \eqref{gauss} by moving away from the optimal jointly Gaussian $(U,X)$. There are several ways to perturb $(U,X)$, we consider first the following case.
We keep $U$ and $V$ independent, but perturb them away from Gaussian's in the following way:
\begin{align}
& p_{U_\e}(u) = g_Q(u) (1+ \e   ( H_3^{[Q]} (u) + \delta H_{4}) ) \label{udef}\\
& p_{V_\e} (v) = g_R(v) (1-\e (H_3^{[R]}(v) - \delta H_{4})) \label{vdef}
\end{align}
with $\e,\delta>0$ small enough. Note that these are valid density functions and that they preserve the first two moments of $U$ and $V$. The reason why we add $\delta H_{4}$, is to ensure that \eqref{c2} is satisfied, but we will see that for our purpose, this can essentially be neglected. Then, using Lemma \ref{hermite2}, the new distribution of $X$ is given by
\begin{align*}
p_X(x) = g_P(x) (1+ \e  \left(\frac{Q}{P}\right)^{\frac{3}{2}} H_3^{[P]} - \e  \left(\frac{R}{P}\right)^{\frac{3}{2}} H_3^{[P]}) +f(\delta)
\end{align*} 
where $f(\delta)=\delta g_P(x)  \e ( \left(\frac{Q}{P}\right)^{\frac{4}{2}} H_4^{[P]} +   \left(\frac{R}{P}\right)^{\frac{4}{2}} H_4^{[P]})$, which tends to zero when $\delta$ tends to zero. Now, by picking $P=2R$, we have
\begin{eqnarray}
p_X(x) &=& g_P(x) + f(\delta). \label{xdis}
\end{eqnarray} 
Hence, by taking $\delta$ arbitrarily small, the distribution of $X$ is arbitrarily close to the Gaussian distribution with variance $P$.
We now want to evaluate how these Hermite perturbations perform, given that we want to maximize \eqref{map}, i.e., 
\begin{align}
&  h(Y_1|U,H) - h(Z_1) + \mu h(Y_2|H) - \mu h(Y_2|U,H). \label{tot}
\end{align}
We wonder if, by moving away from Gaussian distributions, the gain achieved for the term $-h(Y_2|U,H)$ is higher than the loss suffered from the other terms. Using Theorem \ref{hermite2}, Lemma \ref{appro} and Lemma \ref{corr}, we are able to precisely measure this and we get
\begin{align*}
&h(Y_1|U=u,H=h) \\
&=h(g_{h u,v+R h^2} (1- \e  \left( \frac{R h^2}{v+R h^2}\right)^{\frac{3}{2}} H_3^{[hu,v+R h^2]}) ) \\
&= \frac{1}{2} \log 2 \pi e (v+R h^2) - \frac{\e^2}{2}  \left(\frac{R h^2}{v+Rh^2}\right)^{3} + o(\e^2) + o(\delta)
\end{align*}
\begin{align*}
& h(Y_2|U=u,H=h) \\
&= \frac{1}{2} \log 2 \pi e (1+R h^2) - \frac{\e^2}{2}  \left(\frac{R h^2}{1+Rh^2}\right)^{3} + o(\e^2)+ o(\delta)
\end{align*}
and because of \eqref{xdis}
\begin{align*}
h(Y_2|H=h) = \frac{1}{2} \log 2 \pi e (1+P h^2) + o(\e^2) +o(\delta). 
\end{align*}
Therefore, collecting all terms, we find that for $U_\e$ and $V_\e$ defined in \eqref{udef} and \eqref{vdef}, expression \eqref{tot} reduces to
\begin{align}
 & I_G 
  - \frac{\e^2}{2}  \E \left(\frac{R H^2}{v+RH^2}\right)^{3} 
 + \mu  \frac{\e^2}{2}\E  \left(\frac{R H^2}{1+RH^2}\right)^{3} \notag \\
 & +  o(\e^2) +o(\delta) \label{bilan}
\end{align}
where $I_G$ is equal to \eqref{gauss} (and is the mu-rate obtained with Gaussian inputs). Hence, if for some distribution of $H$ and some $v$, we have that 
\begin{align}
   \mu   \E \left(\frac{R H^2}{1+RH^2}\right)^{k}- \E \left(\frac{R H^2}{v+RH^2}\right)^{k} >0,  \label{power3}
\end{align} 	
when $k=3$ and $R$ is optimal for $\mu$, we can take $\e$ and $\delta$ small enough in order to make \eqref{bilan} strictly larger than $I_G$. We have shown how, if verified, inequality \eqref{power3} leads to counter-examples of the Gaussian optimality, but with similar expansions, we would also get counter-examples if the following inequality holds for any power $k$ instead of 3, as long as $k \geq 3$.  
Let us summarize what we obtained: Let $R$ be optimal for $\mu$, which means that
\eqref{optmu} holds if there is only one maxima (not at the boarder). Then, non-Gaussian codes along Hermite's strictly outperforms Gaussian codes, if, for some $k \geq 3$, \eqref{power3} holds. If the maxima is unique, this becomes
$$\frac{\E T(v)^k }{\E T(1)^k} < \frac{\E T(v) }{\E T(1)} $$
where $$T(v)=\frac{R H^2}{v+RH^2}.$$ So we want the Jensen gap of $T(v)$ for the power $k$ to be small enough compared to the Jensen gap of $T(1)$. 

\begin{figure}
\begin{center}
\includegraphics[height=6.6cm,width=7.6cm]{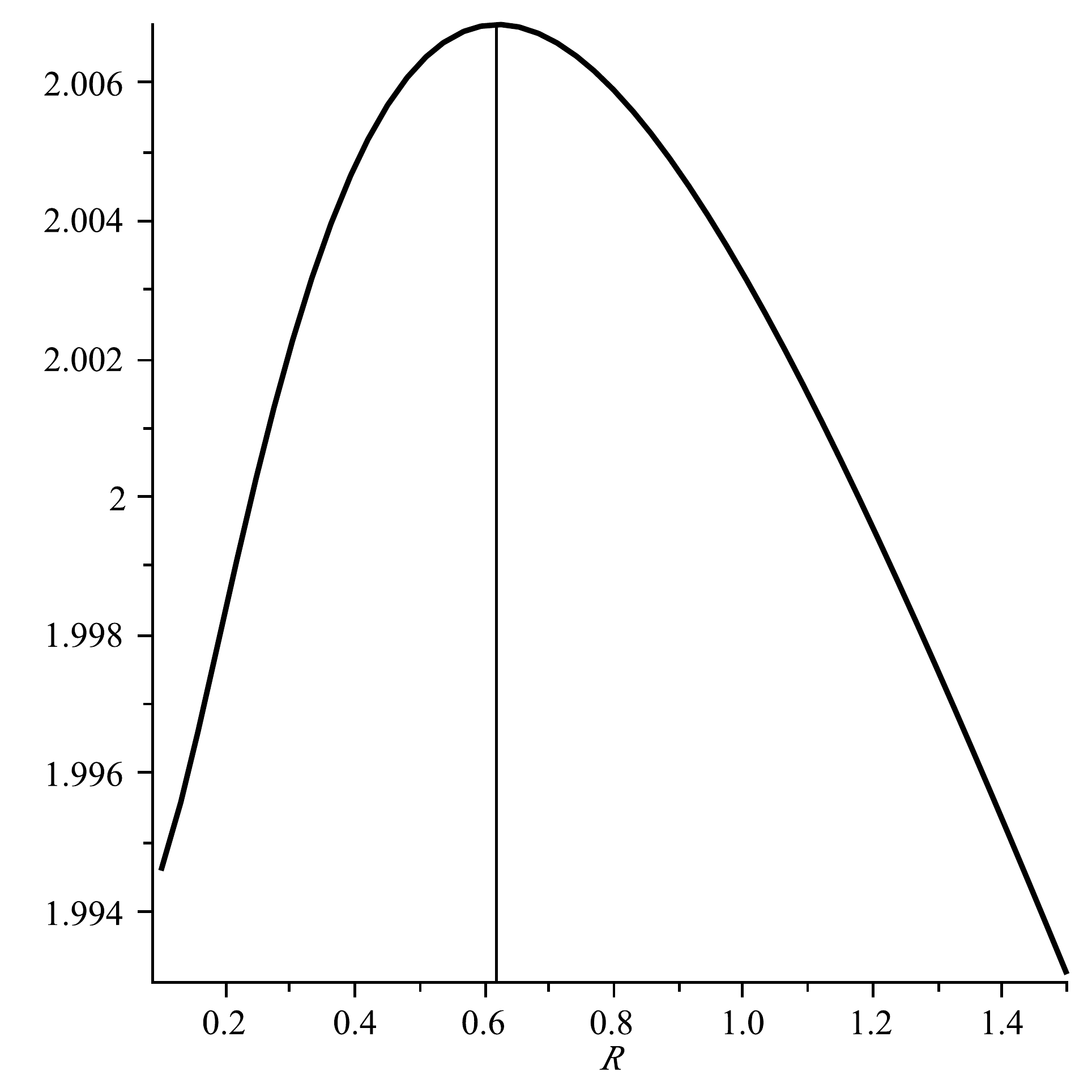} 
\caption{Gaussian mu-rate, i.e., expression \eqref{gauss}, plotted as a function of $R$ for $\mu=5/4$, $v=1/4$, $P=1.24086308$ and $H$ binary $\{1;10\}$. Maxima at $R=0.62043154$.}
\label{maxval}
\end{center}
\end{figure}

\begin{figure}
\begin{center}
\includegraphics[height=6.6cm,width=7.6cm]{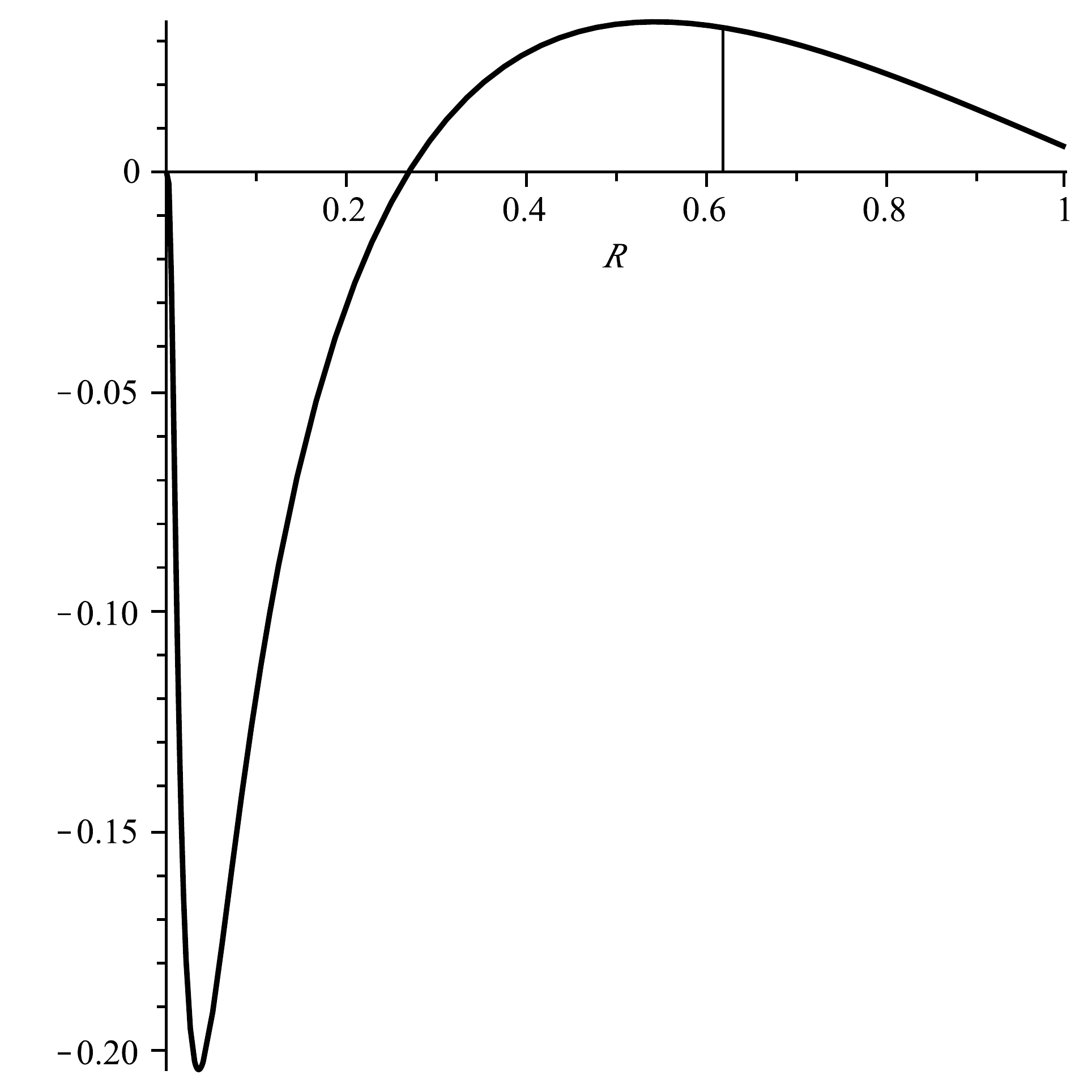} 
\caption{LHS of \eqref{power3} as a function of $R$, for $\mu=5/4$, $v=1/4$, $k=8$ and $H$ binary $\{1;10\}$, positive at $R=0.62043154$.}
\label{ineq}
\end{center}
\end{figure}

We now give an example of a fading distribution for which the above conditions can be verified.
Let $H$ be binary, taking values 1 and 10 with probability half and let $v=1/4$. Let $\mu=5/4$, then for any values of $P$, the maximizer of \eqref{gauss} is at $R=0.62043154$, cf. Figure \ref{maxval}, which corresponds in this case to the unique value of $R$ for which \eqref{optmu} is satisfied.
Hence if $P$ is larger than this value of $R$, there is a corresponding fading BC for which the best Gaussian code splits the power on $U$ and $V$ with $R=0.62043154$ to achieve the best mu-rate with $\mu=5/4$. To fit the counter-examples with the choice of Hermite perturbations made previously, we pick $P=2R$. Finally, for these values of $\mu$ and $R$, \eqref{power3} can be verified for $k =8$, cf. Figure \ref{ineq}, and the corresponding Hermite code (along $H_8$) strictly outperforms any Gaussian codes. 

Note that we can consider other non-Gaussian encoders, such as when $U$ and $V$ are independent with $U$ Gaussian and $V$ non-Gaussian along Hermite's. 
Then, we get the following condition. If for $k \geq 3$ and $R$ optimal for $\mu$, we have
\begin{align}
&   \E \left(\frac{R H^2}{v+RH^2}\right)^{k} \\
 & < \mu  \left[ \E \left(\frac{R H^2}{1+RH^2}\right)^{k} - \left(\frac{R}{P}\right)^k \E \left( \frac{P H^2}{1+PH^2} \right)^k \right], \label{weak}
\end{align} 	
then Gaussian encoders are not optimal. Notice that previous inequality is stronger than the one in \eqref{power3} for fixed values of the parameters. Yet, it can still be verified for valid values of the parameters and there are also codes with $U$ Gaussian and $V$ non-Gaussian that outperform Gaussian codes for some degraded fading BCs. \\

{\it Proof of Theorem \ref{calc}:}\\
Let $\e, \delta >0$ and let $X_1$ and $X_2$ be respectively distributed as $g_p(1+ \e [H_k+\delta H_{4k}])$ and $g_p(1-\e [H_k-\delta H_{4k}])$, where $k \neq 1,2$. We have
\begin{align*}
&I(X_1,X_1+a X_2 + Z_1) = h(X_1+a X_2 + Z_1) - h(a X_2 + Z_1) 
\end{align*}
where $X_k^G$ are independent Gaussian 0-mean and $p$-variance random variables. 
Hence, we need to evaluate the contribution of each divergence appearing in previous expression, in order to know if the perturbations are improving on the Gaussian distributions. Let us first analyze $h(X_1+a X_2 + Z_1)$. The density of $X_1+a X_2 + Z_1$ is given by
\begin{align}
& g_p(1+\e [H_k+\delta H_{4k}]) \star g_{a^2 p}(1-\e [H_k-\delta H_{4k}]) \star g_1, 
 \end{align}
 which, from Theorem \ref{hermite2}, is equal to
\begin{align*}
& g_{p+a^2p+1} (1+  \\
&  \e \{ \left[ \left(\frac{p}{p+a^2p +1} \right)^{\frac{k}{2}}  H_k +\delta \left(\frac{p}{p+a^2p +1} \right)^{2k} H_{4k} \right] \\ 
&-\left[ \left(\frac{a^2p}{p+a^2 p + 1} \right)^{\frac{k}{2}} H_k -\delta \left(\frac{a^2p}{p+a^2 p + 1} \right)^{2k} H_{4k} \right]  \\
& - \e L \}) 
\end{align*}
where $$L =  \frac{g_p [H_k+\delta H_{4k}] \star g_{a^2 p} [H_k-\delta H_{4k}]  \star g_1}{g_{p+a^2p+1}}.$$
Note that each direction in each line of the bracket $\{\cdot\}$ above, including $L$, satisfy \eqref{v1} and \eqref{v2}. Using Lemma \ref{new}, we have
\begin{align}
& L =  \frac{g_{p} [H_k+\delta H_{4k}] \star g_{a^2 p+1} [\left(\frac{a^2 p}{a^2 p+1} \right)^{\frac{k}{2}}H_k- \left(\frac{a^2 p}{a^2 p+1} \right)^{2k} \delta H_{4k}]  }{g_{p+a^2p+1}} \notag \\
& = C_1 H_{2k}^{[p+a^2 p +1]} + C_2 H_{5k}^{[p+a^2 p +1]} + C_3 H_{8k}^{[p+a^2 p +1]}, \label{bug}  
\end{align}
where $C_1,C_2,C_3$ are constants. 
Therefore, the density of $X_1+a X_2 + Z_1$ is a Gaussian $g_{p+a^2p+1}$ perturbed along the direction $H_k$ in the order $\e$ and several $H_{l}$ with $l\geq 2k$ in the order $\e^2$ (and other directions but that have a $\delta$ order). So we can use Lemma \ref{corr} and write 
\begin{align*}
& h(X_1+a X_2 + Z_1) = h(X_1^G+a X_2^G + Z_1)\\
& - D(X_1+a X_2 + Z_1||X_1^G+a X_2^G + Z_1) 
\end{align*}
Using Lemma \ref{appro}, we have 
\begin{align*}
& D(X_1+a X_2 + Z_1||X_1^G+a X_2^G + Z_1) \\
& \eqe  \frac{\e^2}{2} \| \left[ \left(\frac{p}{p+a^2p +1} \right)^{\frac{k}{2}}  H_k +\delta \left(\frac{p}{p+a^2p +1} \right)^{2k} H_{4k} \right] \\ 
&-\left[ \left(\frac{a^2p}{p+a^2 p + 1} \right)^{\frac{k}{2}} H_k -\delta \left(\frac{a^2p}{p+a^2 p + 1} \right)^{2k} H_{4k} \right] \|^2\\
& =  \frac{\e^2}{2} (1 - a^k)^2 \left(\frac{p}{p+a^2 p + 1} \right)^k + \frac{\e^2}{2} o(\delta).
\end{align*}
Hence 
%
\begin{align*}
& h(X_1+a X_2 + Z_1) = h(X_1^G+a X_2^G + Z_1)\\
& -  \frac{\e^2}{2} (1 - a^k)^2 \left(\frac{p}{p+a^2 p + 1} \right)^k 
+ \frac{\e^2}{2} o(\delta).
\end{align*}
Similarly, we get $$D(a X_2 + Z_1||a X_2^G + Z_1) \eqe \frac{\e^2}{2}  \left( \frac{a^2p}{a^2 p + 1} \right)^{k} + \frac{\e^2}{2} o(\delta)$$ and
\begin{align*}
&I(X_1,X_1+a X_2 + Z_1) \eqe I (X_1^G,X_1^G+a X_2^G + Z_1)\\
&+ \frac{\e^2}{2}\left[  \left( \frac{a^2p}{a^2 p + 1} \right)^{k} - (1 - a^k)^2 \left(\frac{p}{p+a^2 p + 1} \right)^k \right] + \frac{\e^2}{2} o(\delta) .
\end{align*}
Finally, we have 
\begin{align*}
&I(X_2,X_2+a X_1 + Z_2) = I(X_1,X_1+a X_2 + Z_1) 
\end{align*}
and
\begin{align*}
&I(X_1,X_1+a X_2 + Z_1)+ I(X_2,X_2+a X_1 + Z_2) \\
& \eqe I (X_1^G,X_1^G+a X_2^G + Z_1) + I (X_2^G,X_2^G+a X_1^G + Z_2) \\& + \e^2 \left[  \left( \frac{a^2p}{a^2 p + 1} \right)^{k} - (1 -a^k)^2 \left(\frac{p}{p+a^2 p + 1} \right)^k \right] + \e^2 o(\delta).
\end{align*}		
Hence, if for some $k \neq 3$ we have 
\begin{align}
 \left(\frac{a^2}{ a^2p+1}\right)^k -  \frac{(a^{k}-1)^2}{(p+a^2p+1)^{k}} >0 \label{thecond} 
\end{align}		
we can improve on the iid Gaussian distributions $g_p$ by using the respective Hermite perturbations.

Now, we could have started with $X_1$ and $X_2$ distributed as $g_p(1+\e b_k \wtil{H}_k)$ and $g_p(1+\e c_k \wtil{H}_k)$, where $\wtil{H}_k$ is defined in \eqref{tild}.
With similar expansions, we would then get that we can improve on the Gaussian distributions if for some some $b_k, c_k$ and $k \neq 1,2$ we have 
\begin{align*}
  &  \left[ \left(\frac{a^2p}{a^2p+1}\right)^k -  \frac{(a^{2}p)^k + p^k}{(p+a^2p+1)^{k}} \right] (b_k^2+c_k^2) \\
  & - 4  \left(\frac{ a p }{p+a^2p+1}\right)^k    b_k c_k >0.
\end{align*}		
But the quadratic function $$(b,c) \in \mR^2 \mapsto \gamma (b^2+c^2)-2 \delta bc,$$
with $\delta>0$, can be made positive if and only if $\gamma +\delta >0$, 
and is made so by taking $b_k=-c_k$. Hence, the initial choice we made about $X_1$ and $X_2$ is optimal. 
Moreover, note that for this distribution of $X_1$ and $X_2$, we could have actually chosen $k=2$ as well. Because, even if Lemma \ref{corr} tells us that we must use correction terms, these correction terms will cancel out when we consider the sum-rate, since $b_k = -c_k$ and since the correction is in $\e$. There is however another problem when using $k=2$, which is that $g_p(1+\e  H_2)$ has a larger second moment than $p$. However, if we use a scheme of block length 2, we can compensate this excess on the first channel use with the second channel use, and because of the symmetry, we can achieve the desired rate. But this is allowed only with synchronization. 
We could also have used perturbations that are mixtures of Hermite's, such as $g_p(1 + \e \sum_k b_k H_k)$. We would then get mixtures of previous equations as our condition. But in the current problem this will not be helpful. Finally, perturbing iid Gaussian inputs in a independent but non i.d. way, i.e., to perturb different components in different Hermite directions, cannot improve on our scheme, from previous arguments. The only option which is not investigated here (but in a work in progress), is to perturb iid Gaussian inputs in a non independent manner. 
Finally, if we work with $k=1$, the proof sees the following modification. In \eqref{bug}, we now have a term in $H_2$. However, even if this term is in the order $\e^2$, we can no longer neglect it, since from Lemma \ref{corr}, a $\e^2 H_2$ term in the direction comes out as a $\frac{\e^2}{\sqrt{2}}$ term in the entropy. 
Hence, we do not get the above condition for $k=1$, but the one obtained by replacing $(a^{k}-1)^2$ with $(a^{2}+1)$, and the condition for positivity can never be fulfilled.  \\

{\it Proof of Proposition \ref{f2}:}\\
From Theorem \ref{calc}, we know that when treating interference as noise and when $F_2(a,p)>0$, it is better to use encoders drawn from the 2 dimensional distributions $X_1^2$ and $X_2^2$, where $(X_1)_1 \sim g_p(1+ \e \wtil{H}_2)$, $(X_2)_1 \sim g_p(1- \e \wtil{H}_2)$, $(X_1)_2 \sim g_p(1- \e \wtil{H}_2)$ and $(X_2)_2 \sim g_p(1+ \e \wtil{H}_2)$, as opposed to using Gaussian distributions. But perturbations in $H_2$ are changing the second moment of the input distribution. Hence, this scheme is mimicing a time-sharing in our local setting. 
Moreover, a direct computation also allows to show that, constraining each user to use Gaussian inputs of arbitrarily block length $n$, with arbitrary covariances having a trace bounded by $nP$, the optimal covariances are $pI_n$ if $F_2(a,p)\leq0$, and otherwise, are given by a time-sharing scheme (cf. Definition \ref{ts} for the definition of a Gaussian time-sharing scheme and covariance matrices). \\

{\it Proof of Proposition \ref{b2}:}\\
Note that when using blind time-sharing, no matter what the delay in the asynchronization of each user is, the users are interfering in $n/4$ channel uses and have each a non-intefering channel in $n/4$ channel uses (the rest of the $n/4$ channel uses are not used by any users). Hence, if the receiver have the knowledge of the asynchronization delay, the following sum-rate can be achieved: $\frac{1}{4}(\log(1+2p)+\log(1+\frac{2p}{1+2a^2 p}))$. And if the delay is unknown to the receivers, the previous sum-rate can surely not be improved on. \\

\begin{figure}
\begin{center}
\includegraphics[scale=.39]{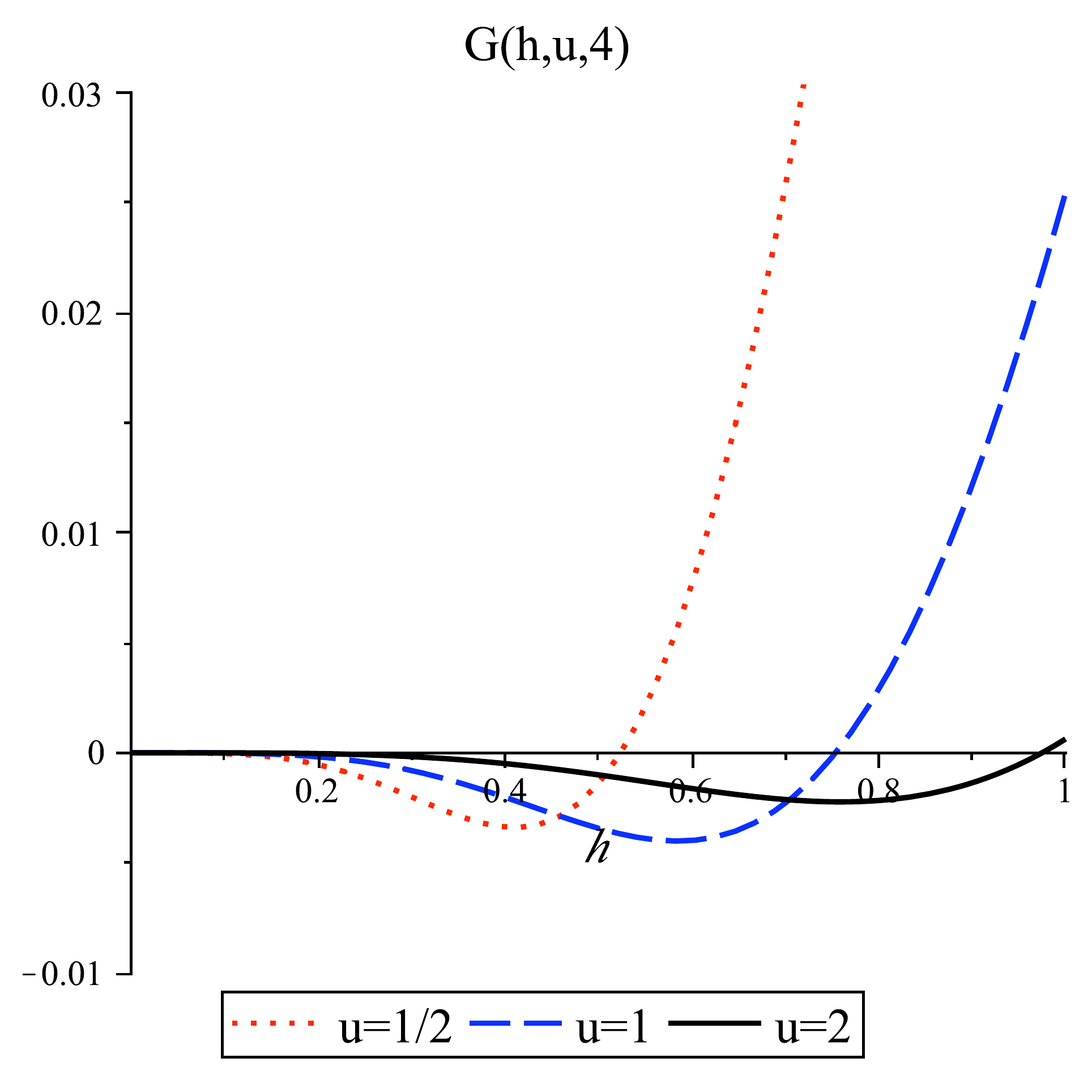} 
\caption{}
\label{slplot}
\end{center}
\end{figure}

{\it Disproof of Conjecture \ref{sl}:}\\
This proof uses similar steps as previous proofs.
Using Lemma \eqref{mean-var}, we express
\begin{align*}
I(X; X + h X_1^G + Z)  \leq I(X; X + h X_1 + Z) .
\end{align*}
as
\begin{align}
&- D( X + h X_1^G + Z || X^G + h X_1^G + Z)  \notag \\
 & \leq - D( X + h X_1 + Z || X^G + h X_1^G + Z) \notag \\
& + D(h X_1 + Z || h X_1^G + Z). \label{tot}
\end{align}
We then pick $X,X_1 \sim g_p(1+ \e \wtil{H}_k)$ and assume that $k$ is even for now.
We then have
\begin{align*}
& X + h X_1^G + Z  \\
& \sim g_p(1+ \e  H_k ) \star g_{ph^2 + v} \\
& = g_{p+ ph^2 + v} (1 + \e (\frac{p}{p+ ph^2 + v})^{k/2} H_k )
\end{align*}
and 
\begin{align*}
&D( X + h X_1^G + Z || X^G + h X_1^G + Z)  \\
& \eqe \frac{\e^2}{2} (\frac{p}{p+ ph^2 + v})^{k}.
\end{align*}
Similarly, 
\begin{align*}
& X + h X_1 + Z  \\
& \sim g_p(1+ \e H_k ) \star g_{p h^2}(1+ H_k ) \star g_v\\
& = g_p(1+ \e  H_k ) \star g_{p h^2+v}(1+ (\frac{ph^2}{ph^2 + v})^{k/2} H_k )\\
& \, g_{p + p h^2+v}(1+ \e (\frac{p}{p + p h^2+v})^{k/2} H_k \\&+ \e (  \frac{ph^2}{p + p h^2+v})^{k/2}  H_k )
\end{align*}
and 
\begin{align*}
&D( X + h X_1 + Z || X^G + h X_1^G + Z)  \\
& \eqe  \frac{\e^2}{2} ( \frac{p}{p + p h^2+v})^{k} (1+h^k)^2 .
\end{align*}
Finally,
\begin{align*}
& h X_1 + Z \sim g_{p h^2  }(1 + \e H_k) \star g_v \\
& = g_{h^2 p +v}(1 + \e (\frac{p h^2}{p h^2 + v})^{k/2} H_k)
\end{align*}
and
\begin{align*}
&D( h X_1 + Z || h X_1^G + Z)  \eqe  \frac{\e^2}{2} (\frac{p h^2}{p h^2 + v})^{k},
\end{align*}
Therefore, \eqref{tot} is given by
\begin{align*}
&- \left(\frac{p}{p+ ph^2 + v}\right)^{k} \\& \leq - \left( \frac{p}{p + p h^2+v}\right)^{k} (1+h^k)^2 + \left(\frac{p h^2}{p h^2 + v}\right)^{k} + o(1)
\end{align*}
and if 
\begin{align}
& \left( \frac{p}{p + p h^2+v}\right)^{k} (1+h^k)^2 - \left(\frac{p}{p+ ph^2 + v}\right)^{k} \notag \\
& - \left(\frac{p h^2}{p h^2 + v}\right)^{k}  >0  \label{plot}
\end{align}
for some $k$ even and greater than 4, we have a counter example to the strong conjecture. Note that, using the same trick as in previous proofs, that is, perturbing along $\wtil{H}_k$ instead of $H_k$, we get that if \eqref{plot} holds for any $k \geq 3$, we have a counter example to the strong conjecture. 
Defining $u:=v/p =1/\text{SNR}$, \eqref{plot} is equivalent to
\begin{align}
&G(h,u,k):= \notag \\
&\frac{(1+h^k)^2}{(1 +  h^2+u)^k}  - \left(\frac{1}{1+ h^2 + u}\right)^{k}   - \left(\frac{ h^2}{ h^2 + u}\right)^{k}  >0 . \label{plot2}
\end{align}
As shown in Figure \ref{slplot}, this can indeed happen.
An interesting observation is that the range where \eqref{plot2} holds is broader when $u$ is larger, i.e., when SNR is smaller. Indeed, when $u=v=0$, which corresponds to dropping the additive noise $Z$, we do not get a counter-example to the conjecture. But in the presence of Gaussian noise, the conjecture does not hold for some distributions of $X,X_1$. The conjecture had been numerically checked with binary inputs at low SNR, and in this regime, it could not be disproved. With the hint described above, we checked numerically the conjecture with binary inputs at high SNR, and there we found counter-examples. 

\section{Discussion}

We have developed a technique to analyze codes drawn from non-Gaussian ensembles using the Hermite polynomials. If the performance of non-Gaussian inputs is usually hard to analyze, we showed how with this tool, it reduces to the analysis of analytic power series. 
This allowed us to show that Gaussian inputs are in general not optimal for degraded fading Gaussian BC, although they might still be optimal for many fading distributions. 
For the IC problem, 
we found that in the asynchronous setting and when treating interference as noise, using non-Gaussian code ensembles ($H_3$ perturbations) can strictly improve on using Gaussian ones, when the interference coefficient is above a given threshold, which significantly improves on the existing threshold (cf. \cite{calvo}). 
We have also recovered the threshold of the moderate regime by using $H_2$ perturbations in the synch setting, showing that this global threshold is reflected in our local setting. We also met mysteriously in our local setting the other global threshold found in \cite{veer,khan,kram}, below which treating interference as noise with iid Gaussian inputs is optimal. It is worth noting that this two global thresholds (moderate regime and noisy interference) are recovered with our tool from a common analytic function. We hope to understand this better with a work in progress.    

The Hermite technique provides not only counter-examples to the optimality of Gaussian inputs but it also gives insight on the competitive situations in Gaussian network problems. For example, in the fading BC problem, the Hermite technique gives a condition on what kind of fading distributions and degradedness (values of $v$) non-Gaussian inputs must be used. 
It also points out that the perturbation in $H_3$ are most effective when carried in an opposite manner for the two users, so as to make the distribution of $X$ close to Gaussian. 

Finally, in a different context, local results could be ``lifted'' to corresponding global results in \cite{lud}. There, the localization is made with respect to the channels and not the input distribution, yet, it would be interesting to compare the local with the global behavior for the current problem too. The fact that we have observed some global results locally, as mentioned previously, gives hope for possible local to global extensions.
A work in progress aims to use our tool beyond the local setting, in particular, by analyzing all sub-Gaussian distributions. Moreover, there are interesting connections between the results developed in this paper and the properties of the Ornstein-Uhlenbeck process. Indeed, some of these properties have already been used in \cite{naor} to solve the long standing entropy monotonicity conjecture, and we are currently investigating these relations from closer.



\section*{Acknowledgment}
The authors would like to thank Tie Liu and Shlomo Shamai for pointing out problems relevant to the application of the proposed tool, Daniel Stroock for discussions on Hermite polynomials, and Emre Telatar for stimulating discussions and helpful comments.   


\end{document}